# Toward the Origins of Binding Energy Shifts and "Satellites" Formation During Plasma-XPS Measurements

J. Trey Diulus,[1] Ashley R. Head,[2] Jorge Anibal Boscoboinik,[2] Carles Corbella Roca,[3] Alexander Tselev,[4] Andrei Kolmakov[3,a]

[1]Nanoscale Device Characterization Division, PML, NIST, Gaithersburg, MD, 20899,
[2]Center for Functional Nanomaterials, Brookhaven National Laboratory, Upton, NY 11973
[3]Materials Science and Engineering Division, MML, NIST, Gaithersburg, MD, 20899
[4]University of Aveiro, 3810-193 Aveiro, Portugal

[a)] Email: andrei.kolmakov@nist.gov

**Abstract**

In-plasma X-ray photoelectron spectroscopy (plasma-XPS) emerges as a powerful platform for real-time, *in situ* chemical analysis under conditions relevant to semiconductor processing and other plasma-enabled technologies. This study investigates the origins of binding energy (BE) shifts and "satellite" peaks formation observed during plasma-XPS measurements across conductive, dielectric, and gas-phase systems. Using a standard laboratory-based ambient pressure XPS apparatus coupled with an alternating current (AC)-driven capacitively coupled plasma source, we show that metastable surface species, such as transient Au oxides, can be detected during plasma exposure, revealing chemical states hardly accessible using conventional ultrahigh vacuum (UHV) XPS. In dielectric samples (e.g., undoped diamond, sapphire), we observe pressure- and plasma-type-dependent BE shifts up to >50 eV, attributed to X-ray-induced and plasma-mediated surface charging. These shifts are mitigated at higher pressures/plasmas or in electronegative plasmas (e.g., $O_2$), the latter due to enhanced charge compensation mechanisms involving slow negative ions. For gas-phase species, AC-plasma excitation leads to spectral broadening and the emergence of "satellite" peaks with a few eV energy separations, linked to oscillating local plasma potentials in the probing volume. These findings highlight the important and complex interplay of plasma parameters, surface charging, and local electric fields in shaping XPS spectra. Overall, plasma-XPS emerges as a critical metrological tool for probing transient surface chemistry, with implications for semiconductor processing, material synthesis, and plasma diagnostics.

# I. Introduction

X-ray photoelectron spectroscopy (XPS) is one of the most surface-sensitive, informative, and well-established electron spectroscopy techniques for surface chemical state analysis, and therefore it is routinely used for wafer analysis in the semiconductor industry. However, due to standard XPS's ultra-high vacuum (UHV) requirements, the wafer fabrication process has to be interrupted, and the sample has to be (in-vacuo) transferred from the plasma reaction chamber for XPS analysis. [1,2] On the other hand, real-time analysis of the chemical composition of the interfaces with sub-monolayer precision is an attractive opportunity for plasma-assisted process control and development in modern semiconductor fabrication technology. These processes usually run from $10^{-1}$ Pa to $10^3$ Pa under a plasma environment and, therefore, are hard to interface with standard UHV surface science analytical equipment. The application of electron spectroscopies such as UV laser-induced electron photoemission for *operando* plasma etching endpoint detection, [3,4] or plasma-enhanced chemical vapor deposition (PECVD) film growth[5] has been reported already in the 1980's. However, the chemical specificity of these techniques was rather limited, which precludes the identification of the process-specific short-lived reactive intermediates. Many of the challenges have been neatly resolved by the Donnelly group using the spinning wall approach[6] where the fast-rotating surface of interest can be analyzed in high vacuum using Auger electron spectroscopy[7] as fast as a millisecond after plasma exposure. The adaptation of this method for planar samples, such as Si-wafers, has yet to be realized. Alternatively, the through-the-membrane XPS method[8,9] has been proposed to probe the interfaces at elevated pressures, but the method's applicability for plasma XPS studies and the longevity of ultrathin membranes in plasmas have yet to be proven.[10,11]

On the other hand, the modern ambient pressure XPS (APXPS, also known as near ambient pressure (NAP) XPS) instrumentation covers the pressure range used in standard plasma applications, and its application space, in this regard, can easily be extended to the plasma environment. Recently we [12] and others[13,14,17] demonstrated that good quality XPS data can be collected during plasma-induced oxidation-reduction reactions, at least under quasi-remote plasma conditions. We also showed the influence of the plasma chamber wall reactions on sample surface chemistry, as well as the ability of the plasma-XPS to analyze the plasma chemistry in the gas phase.[16] Moreover, the electron spectrometer's capacity to measure the electron energy distribution even without X-ray excitation[17] demonstrates the potential of APXPS for plasma diagnostics.

Dielectric materials and coatings dominate real-world samples and devices. Though the charging and charge neutralization mechanisms of poorly conducting samples during XPS, both in vacuum and elevated pressures, are well documented, the charge-induced binding energy (BE) shifts under plasma conditions have not been studied yet. On the other hand, plasma-induced charging and damage of wafers with an insulating layer is a well-known semiconductor fabrication challenge. [18,19] Thus, plasma-XPS metrology can be a useful tool for better diagnostics and mitigation of the latter effects.

Here, in addition to the Au oxidation example, we test plasma-XPS metrology in application to dielectric and gas phase samples. We observe plasma condition-dependent anomalous XPS binding energy shifts due to sample charging and propose mechanisms responsible for this. We also observe and discuss the plasma-induced BE shifts, broadening, and "satellite" formations when measuring plasma gas phase XPS.

## II.  Experimental

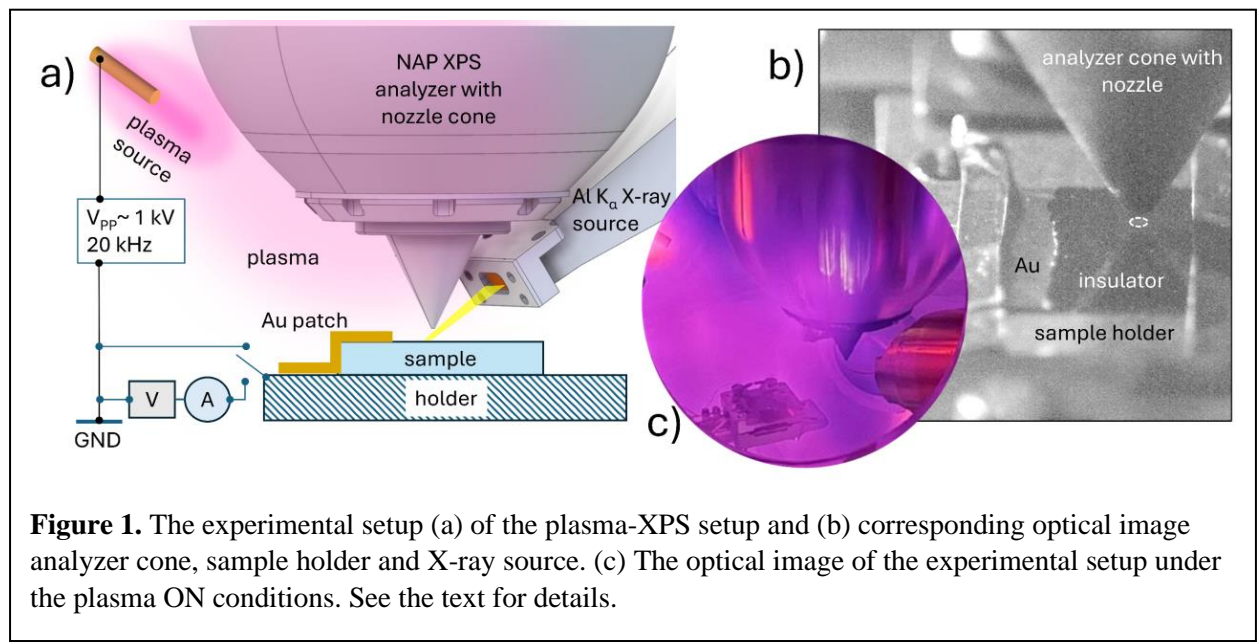

**Figure 1.** The experimental setup (a) of the plasma-XPS setup and (b) corresponding optical image analyzer cone, sample holder and X-ray source. (c) The optical image of the experimental setup under the plasma ON conditions. See the text for details.

Plasma-XPS experiments were conducted at the Center for Functional Nanomaterials at Brookhaven National Laboratory, which houses a lab-based APXPS system. Further details about the experimental setup and its capabilities can be found elsewhere.[12,15,16] The APXPS instrument (Figure 1) features a reaction chamber with a base pressure of <5×10$^{-7}$ Pa and the capability to backfill the chamber with the gas of interest during data collection. The reaction chamber is separated from the multi-stage, differentially pumped electrostatic focusing lens system of the APXPS electron energy analyzer by a 300 μm diameter cone aperture, enabling XPS data collection at pressures up to 200 Pa. A monochromatized Al Kα X-ray source (hν = 1487 eV), focused to a spot size of about 300 μm and fixed at a 30° angle from the sample normal, was used for photoemission excitation. The sample-to-nozzle distance was adjusted to the focal point of the X-rays, which is approximately 600 μm away from the aperture, by optimizing the intensity of a photoemission peak. Gas-phase XPS spectra were collected with the sample retracted centimeters away from the focus of both the X-ray source and analyzer. The sample, an undoped 5 mm × 5 mm × 0.5 mm diamond or 7 mm × 7 mm ×1 mm sapphire dielectric with approximately one-quarter of its surface area coated with a 200 nm Au film, was attached to a sample holder with Ta strips via spot welds and loaded into the APXPS chamber. This Au patch (Figure 1), electrically connected to the sample holder, served as a reference material for XPS spectra. For the plasma-induced oxidation experiment, a polycrystalline 99.95 % 100 μm thick Au foil was used. Prior to oxidation, the foil was first exposed to $H_2$ plasma to remove surface carbon and then annealed to

400 K in UHV and then in $O_2$ to further remove adventitious carbon species. The sample holder can be either grounded or biased and used to monitor the photoemission current or the plasma-induced current. To measure the current, the ground connector on the manipulator was connected to an electrometer. Typically, the photoemission current for a well-focused X-ray source is within the range of 100 pA ± 10 pA, and the measured plasma-induced current is in the range 0 µA to 100 µA, depending on discharge and sample bias parameters (see also SI). The holder can be heated *in situ* to up to approximately 700 K in a gaseous environment.

Plasma was ignited in the APXPS by backfilling the analysis chamber with 5 Pa to 200 Pa of the gas of interest ($H_2$, He, $O_2$, or $N_2$) and applying high AC voltage (22 kHz, up to a one keV peak-to-peak (pk-pk)) to a copper wire driving electrode mounted on a high-voltage feedthrough (Figure 1), located approximately 10 cm from the analyzer's focal point. The high-voltage power supply is set to output from 12 % to 40 % of its nominal power in our experiments, which approximately peaks to up to a few watts of discharge power (see SI). Langmuir probe measurements, plasma potentials, and ion energy distributions were measured for representative plasma conditions in a separate plasma diagnostic vacuum chamber. The exact plasma power numbers (as well as many other plasma parameters such as plasma potential, floating potential, electron or ion densities, and temperatures) strongly depend on the gas type and pressure used, but for the purposes of this paper, we will refer to the percentage power output of the current setup since the true discharge power scales with those. Some of the results reported below use the pressure excursions under the plasma ON condition. It is important to note that due to the inherent entanglement of many plasma parameters, the discharge emission intensity and its spatial distribution (see SI) respond sensitively to gas pressure changes, which results in significant alteration of the ions, electrons, photons, and neutrals fluxes at the sample surface.

C 1s and Au 4f spectra were collected with a 20 eV pass energy, a 250 ms dwell time, a 50 meV step size, and multiple sweeps for improved signal-to-noise ratio. For data collected at elevated pressure, the number of sweeps was doubled. XPS spectra were collected as a function of gas pressure, plasma power, and composition, temperature, and sampling point location.

### III. RESULTS AND DISCUSSION

#### A. Peak "satellites" observed on well-conducting samples

The plasma-induced chemical shifts observed thus far during operando plasma-XPS[12-14, 16, 17] closely resemble those measured in previous reduction, oxidation, or etching experiments conducted using the "before-and-after" approach. The value of plasma-XPS metrology compared to its standard UHV before-and-after realization is in accessing metastable interfaces, transient or short-living reactive intermediates that vanish immediately after plasma termination. An industrially important example of the metastable interface is Au oxide, which can be formed by exposing a clean Au surface to oxygen under UV light, ozone, or oxygen plasma, but is unstable in UHV and at elevated temperatures above 400 K [20-26].

Figure 2a presents a series of XPS spectra of gold collected under different plasma conditions. The bottom spectrum is taken at 200 Pa, showing a typical Au 4f XPS peak. The middle spectrum is collected at 7 Pa under oxygen plasma conditions, revealing an appearance of an additional peak

(shoulder) indicative of a metastable gold oxide formation. The top spectrum is obtained under UHV conditions after plasma exposure and does not show the Au-O band, as the oxide did not remain after plasma quenching.

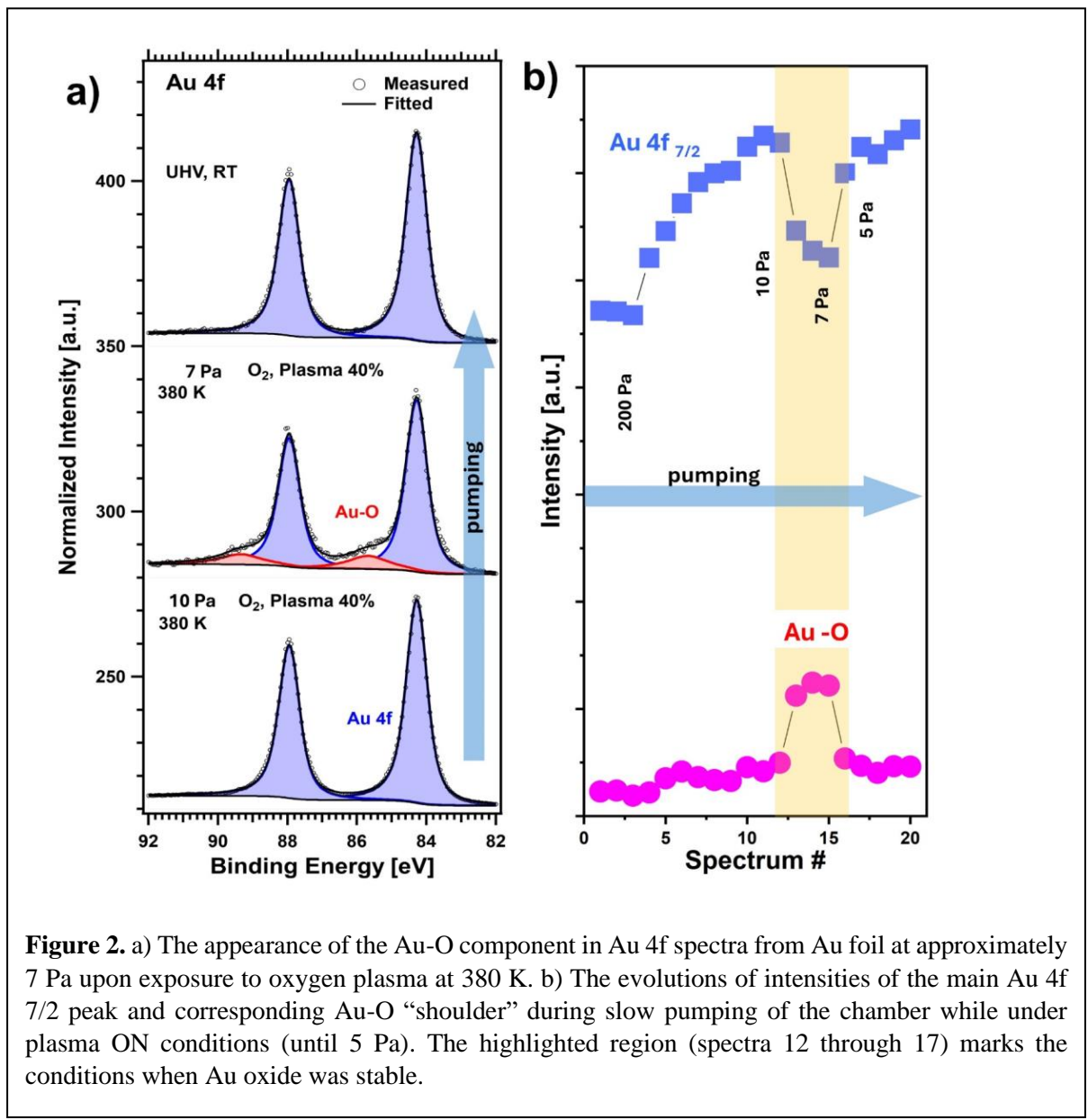

**Figure 2.** a) The appearance of the Au-O component in Au 4f spectra from Au foil at approximately 7 Pa upon exposure to oxygen plasma at 380 K. b) The evolutions of intensities of the main Au 4f 7/2 peak and corresponding Au-O "shoulder" during slow pumping of the chamber while under plasma ON conditions (until 5 Pa). The highlighted region (spectra 12 through 17) marks the conditions when Au oxide was stable.

Figure 2b summarizes 20 sequential spectra collected under plasma ON conditions as the oxygen pressure is gradually reduced from 200 Pa down to ultra-high vacuum. The top curve shows the intensity of the main metallic Au $4f_{7/2}$ peak, which increases as the pressure decreases due to reduced attenuation of photoelectrons by the background gas. The bottom curve indicates the intensity in the "shoulder" region of the expected gold oxide "satellite" around 85.5 eV BE, which grows suddenly within the 10 Pa to 5 Pa pressure range and then disappears as the pressure is further reduced. Note that the intensity of the main Au $4f_{7/2}$ peak is reduced concomitantly at this particular pressure range. We attribute this effect to formation of the Au-oxide at the surface and concomitant attenuation of the Au $4f_{7/2}$ photoelectrons by the newly formed oxide layer. This experiment corroborates with the prior observations[20-26] of the metastability of the gold oxide under high vacuum conditions and elevated temperatures. In our case, the narrow window of the plasma parameters near the sample exists during the pressure decrease, where the dynamic equilibrium of the incoming oxygen radicals responsible for Au oxide formation overwhelms the $O_2$ desorption rate, and as a result Au-oxide can be detected at the surface.

### B. BE shifts observed on insulator samples

We have studied the plasma-induced charge accumulation and neutralization during XPS

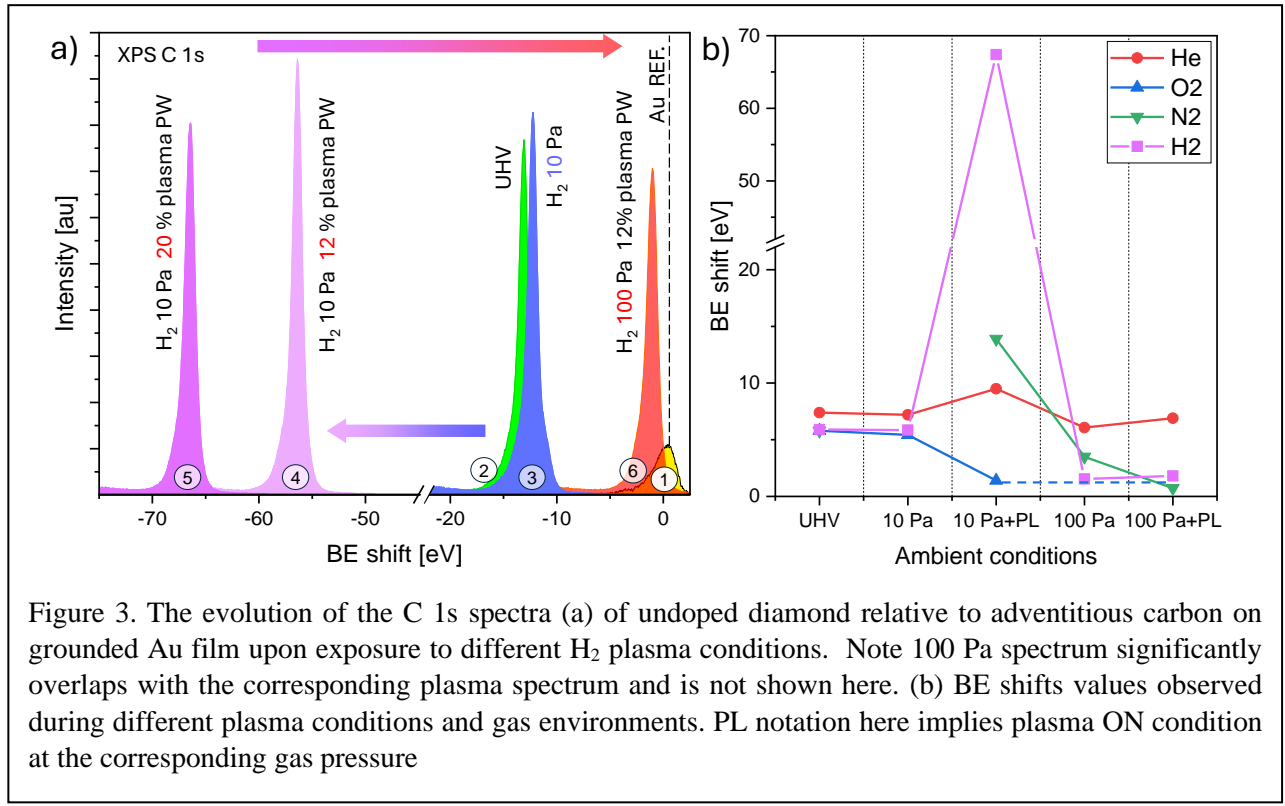

Figure 3. The evolution of the C 1s spectra (a) of undoped diamond relative to adventitious carbon on grounded Au film upon exposure to different $H_2$ plasma conditions. Note 100 Pa spectrum significantly overlaps with the corresponding plasma spectrum and is not shown here. (b) BE shifts values observed during different plasma conditions and gas environments. PL notation here implies plasma ON condition at the corresponding gas pressure

measurements. The standard experiment routine includes sequential XPS on the reference Au patch area (spectrum 1 in the Figure 3a) and dielectric sample center under UHV conditions (spectrum 2), followed by similar measurements at 10 Pa of $H_2$ (spectrum 3). After that, XPS under 12 % and 20 % of the output plasma power were collected (spectra 4 and 5 correspondingly). The plasma was then turned off, and the pressure in the chamber was increased to 100 Pa of the gas of interest. XPS spectra were collected again (overlaps with spectrum 6, not shown), followed by corresponding in-plasma measurements (spectrum 6). The chamber was then evacuated to UHV, and reference spectra from the grounded Au patch were checked again.

Figure 3a depicts the typical evolution of the C 1s XPS peak position collected from the middle of the single crystal undoped diamond under the different hydrogen gas and plasma environments.

As can be seen, the C 1s spectrum collected in a vacuum is shifted approximately 14 eV toward higher BE in our experimental settings, which reflects the local positive charging of the X-ray irradiated spot of the dielectric up to the surface potential value until outgoing photoemission, Auger, and secondary electron fluxes become compensated through multiple charge neutralization channels [27, 28] such as sample surface/bulk leakage current, photoemission from surrounding parts, and recapture of the low-energy secondary electrons. Note that leakage current in dielectrics and diamond in particular also includes X-ray-induced photocurrent.

The admission of 10 Pa of hydrogen induces a slight reverse shift of the C 1s peak to the lower binding energy due to the appearance of an additional charge neutralization channel via ionization of the nearby ambient gas by the primary X-ray beam and by outgoing photo-emitted and energetic secondary electrons.[29] The increase in pressure to 100 Pa makes this charge neutralization mechanism dominate and leads to nearly complete surface charge compensation (see below), which has also been observed in multiple APXPS reports. [30-34]

The ignition of the plasma at 10 Pa results in a significant (> 50 eV) additional shift toward higher binding energy. Increasing the plasma power to 20 % enhances the effect even further (> 65 V, see Figure 3a, compare spectra 4 and 5). This drastic plasma-induced BE shift can be almost completely nullified if the hydrogen pressure is raised to 100 Pa while still under the plasma ON condition. Note that in our settings, the absolute values of the BE shift strongly depend on the experiment geometry, plasma type/conditions, sample type (diamond or sapphire), temperature, and even irradiation spot location on the same sample. However, the pressure results are reproducible for the same sample with an error bar of approximately 3 eV provided all other experimental parameters are locked.

Figure 3b summarizes the BE shift results for undoped diamonds under different conditions and plasma gas types. Overall, the trend for all studied plasmas (He, $H_2$, $N_2$) and also for the sapphire sample (see SI) is qualitatively the same: lower pressure 10 Pa plasmas cause significant plasma-induced surface positive charging, while a pressure increases to 100 Pa, with or without plasma, significantly compensates surface charge. The oxygen plasma, however, deviates significantly from the above trend: no plasma-induced charging can be observed, and instead, even 10 Pa $O_2$ plasma compensates for surface positive charge almost completely.

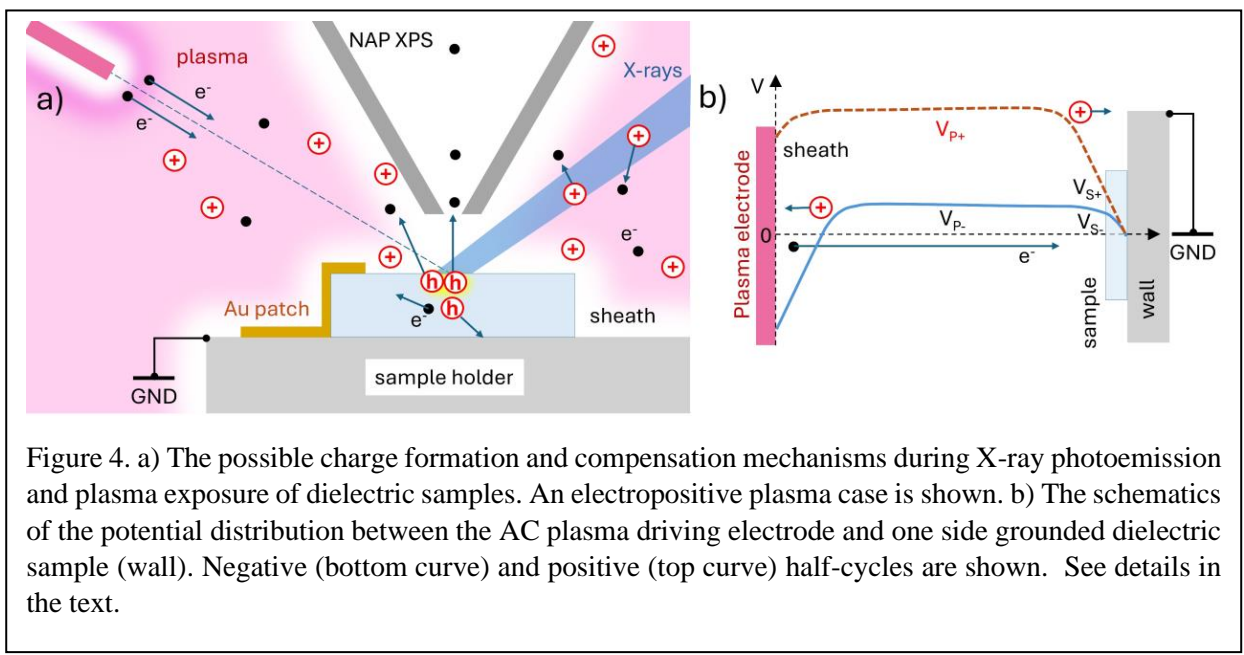

Figure 4. a) The possible charge formation and compensation mechanisms during X-ray photoemission and plasma exposure of dielectric samples. An electropositive plasma case is shown. b) The schematics of the potential distribution between the AC plasma driving electrode and one side grounded dielectric sample (wall). Negative (bottom curve) and positive (top curve) half-cycles are shown. See details in the text.

A few mechanisms can be invoked to explain qualitatively the appreciable (10 V to 100 V) additional to X-ray positive charging of the dielectric sample in a plasma environment in our experimental settings (Figure 4 a, b).

The (vacuum) ultraviolet ((V)UV) plasma emission can constitute a noticeable partition of the energy flux to the sample[35, 36], induce the measurable photoemission from the sample and therefore its charging, if not compensated. However, the observed values of plasma-induced BE shifts significantly exceed the most extreme UV radiation energies of the used plasmas, and therefore, the (V)UV photoemission-induced charging mechanism can be ruled out. On the contrary, VUV light sources are often used in industry for charge compensation in insulating samples in vacuum via (V)UV-induced photoconductivity (leakage current) and secondary electron emission (and/or residual gas ionization) of the sample surroundings[37].

The alternative explanation relies on the fact that plasma potential is always positive with respect to the grounded walls and sample holder, independent of the positive or negative driving potential of the driving electrode (see diagram in Figure 4b). Moreover, the estimated plasma frequencies in our settings are at least an order of magnitude larger compared to the driving frequency (see SI, section F), and we may safely conclude that, to a first approximation, our plasma can be treated as a plasma of a DC glow discharge, where the plasma-generating electrode acts as a cathode or anode. The thin dielectric sample rests on the other grounded electrode, and its surface potential is floating. When the plasma-generating electrode acts as the anode, the floating potential of the plasma bulk ("positive column") $V_{P+}$ is at the highest positive potential with respect to the chamber wall and sample holder (top curve in the Fig. 4b). The maximum value of this potential is even higher than the amplitude of the applied voltage (which is up to 1 kV). At the same time, on the chamber wall, there is a thick cathode sheath with a distributed space charge whose main function is to provide high-energy ions for the generation of secondary electrons from the metal cathode through bombardment to maintain gas ionization and passage of the electric current through the gas. The surface of a dielectric sample resting on such a cathode is generally charged positively due to secondary electron emission as a result of ionic bombardment. The steady-state potential on the dielectric surface ($V_{S+}$) will acquire a value necessary to stop ionic bombardment and secondary electron emission since almost no current flows through the dielectric. On the other half of a cycle, when the plasma-generating electrode acts as the cathode (bottom curve in Fig. 4b), the floating potential of the plasma bulk ($V_{P-}$) is still positive, however, relatively small, above the zero potential of the chamber wall acting as the anode. Since the charge relaxation time of the dielectric surface is usually much longer than the period of the voltage oscillations, the average potential acquired by the dielectric surface is positive with a steady state value depending on the properties of the plasma sheath, which are, in turn, function of type of the gas, gas pressure, value of the applied voltage, and the current density in the plasma. It can be expected that at 10 Pa, the positive potential on the dielectric surface will be highest for the lowest resistance plasma, which is hydrogen in our case, and indeed has been observed. Overall, the surface of the dielectric becomes charged to the local floating plasma potential, to be in a dynamic equilibrium with the plasma. Measurement of the local plasma potential near the dielectric surface with the use of Langmuir probe yielded values comparable to the BE peak shifts observed in the XPS measurement. It's important to note that no significant broadening of the XPS peak's Full Width at Half Maximum (FWHM) is seen, suggesting the potential is not fluctuating nor is the sample differentially charging across the probed spot.

Another factor favoring the formation of the high positive potential at the sample surface is ballistic high-energy electrons emitted from the driving electrode during its cathode half-cycle. Indeed, at low pressure (10 Pa), the mean free path for a few hundred eV electrons is in the order of ten centimeters, and a significant fraction of the electrons generated at a cathode can gain up to 1 keV energy before impinging on the sample (anode). Similarly, at the most positive potential at the driving electrode, the significant fraction of ions impinges on the sample with $V_{P+}$ potential (Fig. 4b). Given the fact that the dielectric samples have secondary electron yield well above 1 for few hundreds eV primary electrons but significantly low ion-induced secondary electron emission coefficients, the above mechanism will sponsor a positive potential accumulation at the dielectric surface with its steady state value depending on secondary electron emission, leakage and (photo-)currents.

At high pressure, the current density is generally unevenly distributed over the electrode's surface, and correspondingly, the sheath structure and plasma parameters in the sheath are not even along the electrode surface. If the current density in the plasma around the dielectric is small, which is most likely true for our experiments, the dielectric surface will be near the potential of the sample holder. Similarly, with the pressure increasing to 100 Pa, all ballistic mechanisms vanish because of the multiple scattering effects, and the sample surface charge becomes effectively neutralized

The unusual behavior of the surface charge under oxygen plasma (see Fig.3 b) can also be explained in the same terms, taking into consideration the plasma's electronegativity. Even at 10 Pa, the number of free electrons is significantly reduced in favor of the formation of negative ions. These charged species effectively neutralize the surface charge during any part of the AC cycle.

Another alternative explanation also relies on the positive plasma potential around the sample as a major factor responsible for the observed surface charging. However, different from the above scenario, the major ionization source remains X-ray excitation. At lower plasma pressure, the presence of the positive plasma potential coupled with a large electron mean free path reduces the recapture of photoemitted and slow secondary electrons, which leads to observed positive sample charging. As in the prior scenario, the increase in pressure leads to an increase in the electron scattering phenomena and near-surface gas ionization, resulting in effective charge neutralization.

Overall, the observed dielectric sample charging is a complex phenomenon that can be explained via multiple mechanisms. Due to the inherent entanglement of multiple plasma parameters, a more detailed and thorough study (e.g., plasma frequency dependence) is required to discriminate between them.

### C. Gas phase BE shifts

We also observed the dramatic effect of plasma on X-ray photoemission from the gaseous species. To explore the influence of the plasma conditions on gas phase XPS, the gas phase spectra with plasma ON and OFF APXPS data were collected sequentially with the sample holder retracted from the analyzer/X-ray source focal spot to avoid ground and/or sample charging interference. Argon plasma was used as a model system. It was found that gas pressure and plasma excitation parameters have a dramatic effect on the intensity, broadness, and binding energy (BE) positions of the XPS peaks.

Figure 5 depicts Ar 2p, 3s, and 3p spectra collected with and without 10 % AC (1 kV pk-pk, 22 kHz) 500 Pa plasma in our experiment geometry settings. Table I summarizes the peak fitting parameters.

The plasma ignition results in:

(i) Appreciable reduction of the intensity of the main peaks (note the scaling ratio)
(ii) Slight shift of the main peaks position toward the higher BE
(iii) Slight broadening of the main peaks and
(iv) The most pronounced effect is the appearance of prominent, broad "satellite" peaks at higher binding energies.

An important observation is that the energy shift between the "satellites" and corresponding main peaks remains constant across the entire XPS spectrum (approximately 2 eV in our case), including Auger band(s) (compare the peak-satellite shifts in Ar 2p, 3s, 2p spectra). Moreover, the cumulative photoelectron yield calculated as the total area of the main peak (when the plasma is OFF) and the main peak plus corresponding "satellite" (when the plasma is ON) remains nearly the same. The latter indicates that the electrons' transmission to the analyzer lens is not significantly affected by our plasma conditions, and instead, the partitioning of the electrons from the main peak to the "satellite" takes place.

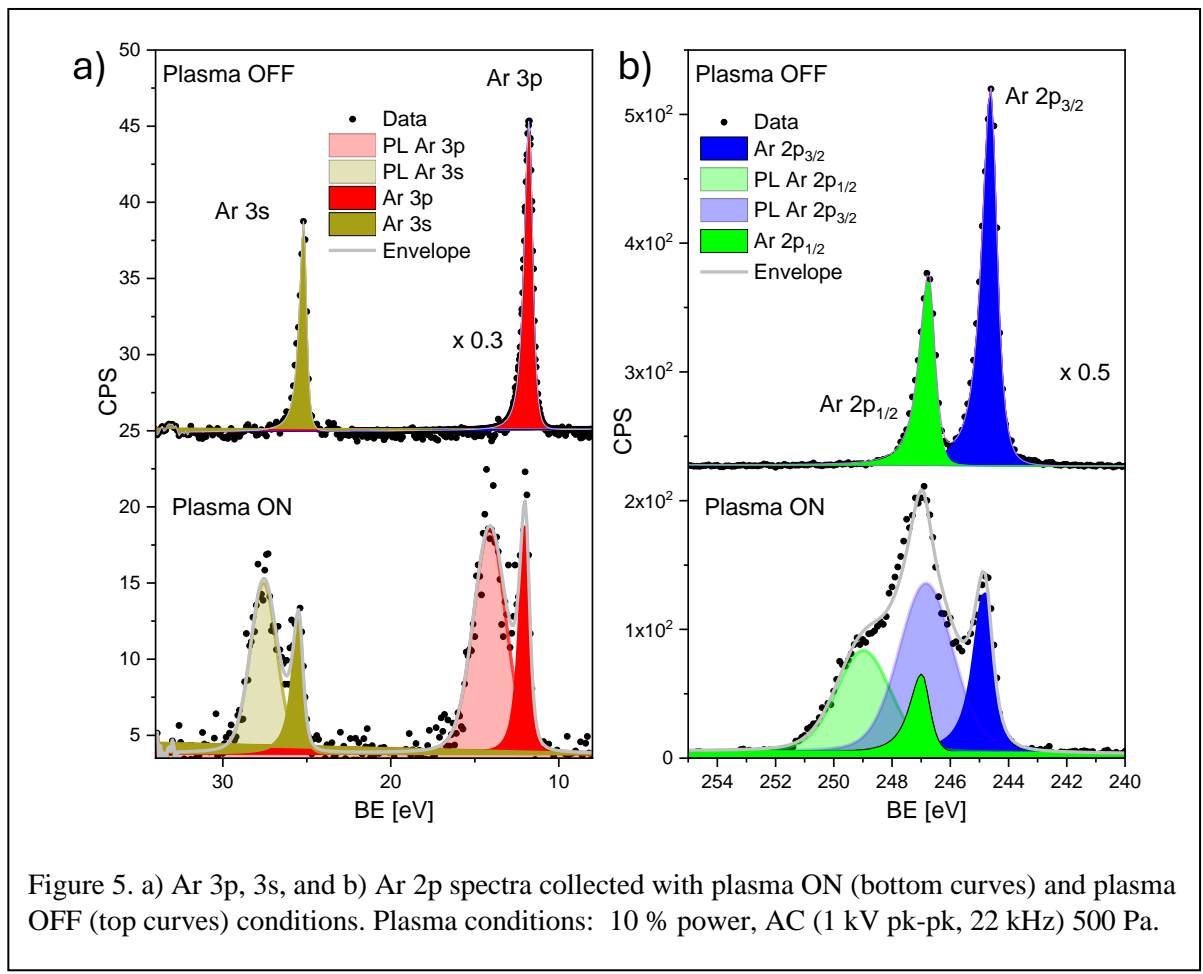

Figure 5. a) Ar 3p, 3s, and b) Ar 2p spectra collected with plasma ON (bottom curves) and plasma OFF (top curves) conditions. Plasma conditions: 10 % power, AC (1 kV pk-pk, 22 kHz) 500 Pa.

The observed effects can be explained by the presence of relatively small (a few electron volt) and oscillating local plasma potential in the X-ray probed volume near the analyzer's focal point (similar to the sample floating potential $V_{S+/-}$ in Figure 4b). It always has a positive value with respect to the ground independently of negative or positive oscillation half-cycle, and its value depends on multiple parameters (geometry and areal ratio between driving electrode and chamber walls, including the analyzer cone, pressure, gas type, driving voltage, etc.)[38]. This explains why the spectra intensity drops upon plasma ignition in favor of the corresponding "satellite" peak, and "satellite" BE is shifted to higher binding energy. Note, this plasma potential oscillates around the mean value with the frequency of the driving source, 22 kHz. This leads to observed significant "satellite" lines broadening, which we see in all the spectra.

**Table I:** Major Ar 2p, 3p, and 3s XPS peak parameters retrieved after the fitting without plasma (left half) and with (shadowed half) plasma ON. Note that despite the BE positions of the main peaks can be measured and fitted with uncertainty of approximately 50 mV in this study, the "satellite's" intensity, BE shifts, peaks broadness sensitively depend on geometry of the experimental setup and plasma parameters.

| Peak assignment | BE [eV] 500 Pa Ar | FWHM [eV] 500 Pa Ar | Area under the peak [CPSxeV] | BE [eV] at 500 Pa Ar + 10 % plasma | FWHM [eV] 500 Pa Ar + 10 % plasma | Area |
|---|---|---|---|---|---|---|
| Ar 3p | 11.8 | 0.5 | 52.2 | 12.0 | 0.6 | 13.8 |
| Ar 3p sat | - | - | - | 14.1 $\Delta$=2.1 | 2.2 | 37.7 |
| Ar 3s | 25.2 | 0.4 | 32.4 $\Sigma$=84.6 | 25.5 | 0.6 | 8.6 |
| Ar 3s sat | - | - | | 27.6 $\Delta$=2.1 | 1.8 | 23.4 $\Sigma$=83.5 |
| Ar 2p3/2 | 244.6 | 0.5 | 480 | 244.8 | 0.7 | 127.8 |
| Ar 2p3/2 sat | - | - | - | 246.8 $\Delta$=1.97 | 2.2 | 332.2 |
| Ar 2p1/2 | 246.8 | 0.5 | 240 $\Sigma$=720 | 247.0 | 0.7 | 63.9 |
| Ar 2p1/2 sat | | | | 249.0 $\Delta$=2.0 | 2.2 | 199.4 $\Sigma$=723.3 |

Such XPS peaks splitting/broadening are commonly observed in AC-biased substates where excitation frequency is comparable or higher than the spectrum acquisition rate (see, for example, ref. [39] and references therein).

## IV. SUMMARY AND CONCLUSIONS

The observed formation and decay of the Au-oxide at the surface of Au foil during oxygen plasma exposure demonstrates the unique capability of plasma-XPS metrology to detect transient chemical states that might not be visible in conventional UHV XPS measurements. The latter is a valuable metrological advancement highly relevant to current needs in semiconductor fabrication, where instant chemical composition of the interfaces is an imperative for the process control.

We also checked the plasma's ability to neutralize the X-ray-induced charging of practically important poorly conducting samples. This initial BE shift in vacuum is routinely observed during XPS on dielectric samples and is due to the spot's local positive charging by X-ray-induced photoemission, Auger processes, and secondary electron emission. The degree of the BE shift under steady-state conditions is defined by the equilibrium between the photoemitted electron flux and charge compensation electron/hole flows in the solid. In a vacuum, the latter is defined by surface and bulk conductance, including photoconductance. Adding the gas to the chamber leads to the appearance of the new charge compensation channel via ionization of the gas in front of the sample by X-rays and outgoing secondary or photoemitted electrons.

We observed that the plasma environment can lead to both: charge neutralization and significant positive charging, depending on (i) pressure and (ii) whether electronegative or electropositive low-pressure plasma is used. At low pressure, the positive plasma potential around the sample and large electron mean free path promote photoemission and reduce the recapture of slow secondary electrons, which leads to observed positive sample charging. The increase in pressure leads to an increase in the electron scattering phenomena and, following effective charge neutralization. In electronegative plasmas, like oxygen in our case, the effective formation and capture of slow negative ions promotes neutralization of the positive charge, even at low pressure.

We also show that the gas phase XPS spectra can be drastically affected by the AC plasma, depending on the plasma conditions. Plasma-induced "satellites" with few eV BE shifts and large peaks broadening appear due to the presence of the positive oscillating plasma potential at the probing volume. To verify the aforementioned proposed hypothesis, more systematic comparative studies under DC and RF plasmas are required.

Overall, plasma XPS metrology is a new, powerful analytical capability that excels in measuring metastable surface states that are impossible or difficult to detect under traditional ultra-high vacuum conditions using before-and-after approach. It's particularly valuable in fields like semiconductor fabrication, biomedical treatments, aerospace applications, and new materials synthesis. Ongoing research into the effects of various electromagnetic fields present in plasma on the formation of XPS spectra will further refine our ability to interpret these complex surface chemistries and can contribute to plasma diagnostic metrologies.

## ACKNOWLEDGMENTS AND DISCLAIMERS

This work was performed with funding from the CHIPS Metrology Program, part of CHIPS for America, National Institute of Standards and Technology, U.S. Department of Commerce. CHIPS for America has financially supported this work through the "Multiscale Modeling and Validation of Semiconductor Materials and Devices project". This research used the Proximal Probes Facility of the Center for Functional Nanomaterials (CFN), which is a U.S. Department of Energy Office of Science User Facility, at Brookhaven National Laboratory under Contract No. DE-SC0012704. The authors are thankful to Dr. Conan Weiland and Dr. Sujitra Pookpanratana (all at NIST) for careful reading of the manuscript and useful suggestions.

## AUTHOR DECLARATIONS

The authors have no conflicts to disclose.

## AUTHOR CONTRIBUTIONS

J. Trey Diulus: Data collection and analysis (lead); Investigation (lead); Writing (supporting). Ashley Head: Data collection (lead), Formal analysis (supporting); Investigation (supporting);

Writing (supporting); Jorge Anibal Boscoboinik: Data collection (lead), Formal analysis (supporting); Investigation (supporting); Writing (supporting), Carles Corbella Roca: Data collection (lead); Investigation (supporting); Writing (supporting). Alexander Tselev: Investigation (supporting); Writing (supporting). Andrei Kolmakov: Conceptualization (lead); Data analysis (lead); Investigation (equal); Supervision (lead); Writing (lead).

DATA AVAILABILITY

The data that support the findings of this study are available from the corresponding author upon reasonable request.

# Toward the Origins of Binding Energy Shifts and "Satellites" Formation During Plasma-XPS Measurements

J. Trey Diulus,[1] Ashley R. Head,[2] Jorge Anibal Boscoboinik,[2] Carles Corbella Roca,[3] Alexander Tselev,[4] Andrei Kolmakov[3,a]

[1]Nanoscale Device Characterization Division, PML, NIST, Gaithersburg, MD, 20899,

[2]Center for Functional Nanomaterials, Brookhaven National Laboratory, Upton, NY 11973

[3]Materials Measurement Science Division, MML, NIST, Gaithersburg, MD, 20899 and Department Chemistry & Biochemistry, University of Maryland, College Park, Maryland 20742

[4]University of Aveiro, 3810-193 Aveiro, Portugal

a) Email: andrei.kolmakov@nist.gov

## A. Optical diagnostics and plasma-induced DC current measurements

Plasma emission spectra and plasma-induced current on the sample holder during plasma exposure have been collected routinely in the main APXPS chamber.

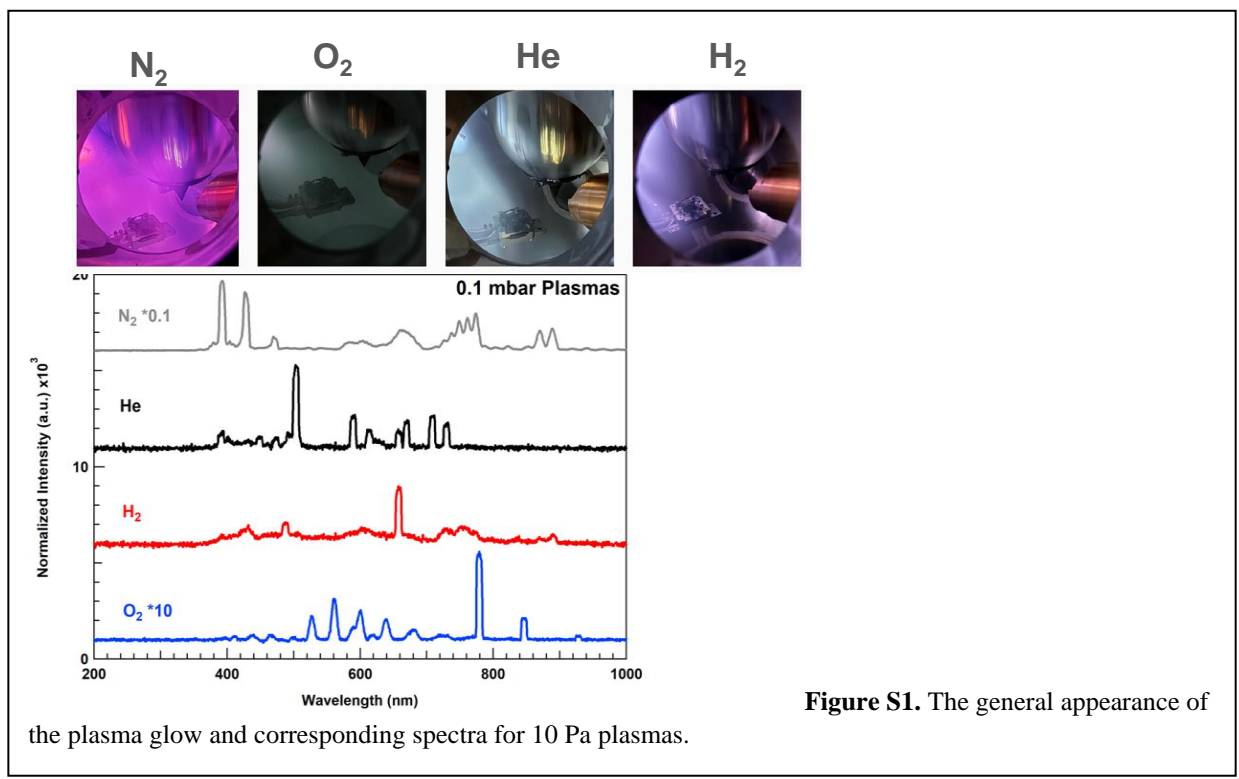

**Figure S1.** The general appearance of the plasma glow and corresponding spectra for 10 Pa plasmas.

Figure S1 depicts the visual appearance (top row) and corresponding low-resolution fast (10 sec integration time) emission spectra (bottom row) of the 10 Pa plasma discharges collected through the viewport borosilicate glass window from the near sample position. The recorded spectra are typical for low density cold plasmas, and in the case of N$_2$, is dominated by First Negative (N$_2^+$) (approximately 390 nm and 430 nm) and First Positive System (500 nm-900 nm) emission bands; He plasma spectrum consists mostly of HeI atomic lines (incl. 388 nm, 502 nm, 588 nm, 706 nm etc) corresponding to the electronic transitions within neutral helium atom. H$_2$ plasma spectrum shows typical Balmer atomic H$_{\alpha,\beta,\gamma}$ lines (657 nm, 486 nm, 434 nm) and molecular Fulcher (550 nm-650 nm, 700 nm-800 nm) bands. The oxygen plasma spectrum is dominated by 777 nm and 845 nm atomic lines and O$_2^+$ molecular 526 nm, 560 nm, 600 nm, and 640 nm lines. No obvious gas contamination can be observed.

## B. Plasma density distribution and plasma-induced currents during pressure change

Due to the inherent entanglement of many plasma parameters, the discharge emission intensity is highly inhomogeneous along the sample-to-source distance, and its spatial distribution responds sensitively to gas pressure changes, which may result in significant alteration of the ions, electrons, photons, and neutrals fluxes at the sample surface. Figure S2 left panels illustrate how the discharge glow concentrates around the driving electrode upon pressure increase. As observed in all the gases considered here, the glow was practically extended to the whole chamber at low pressures, while it tended to contract near the plasma source as pressure increased (Fig. S2 left panel). The same pattern was observed in the reactor where plasma-XPS experiments were conducted. This trend corroborates the plasma-induced current evolution with pressure (right panel in Fig.S2). With the pressure increase, the collected current on the +1 V biased sample holder gradually decreases, indicating the overall reduction of the electron flux toward the sample holder. This is not the case for electronegative O$_2$ plasma, where the collected current changes polarity and saturates at elevated pressure. The spikes and steps in the curves are due to the manual control of the leak valve.

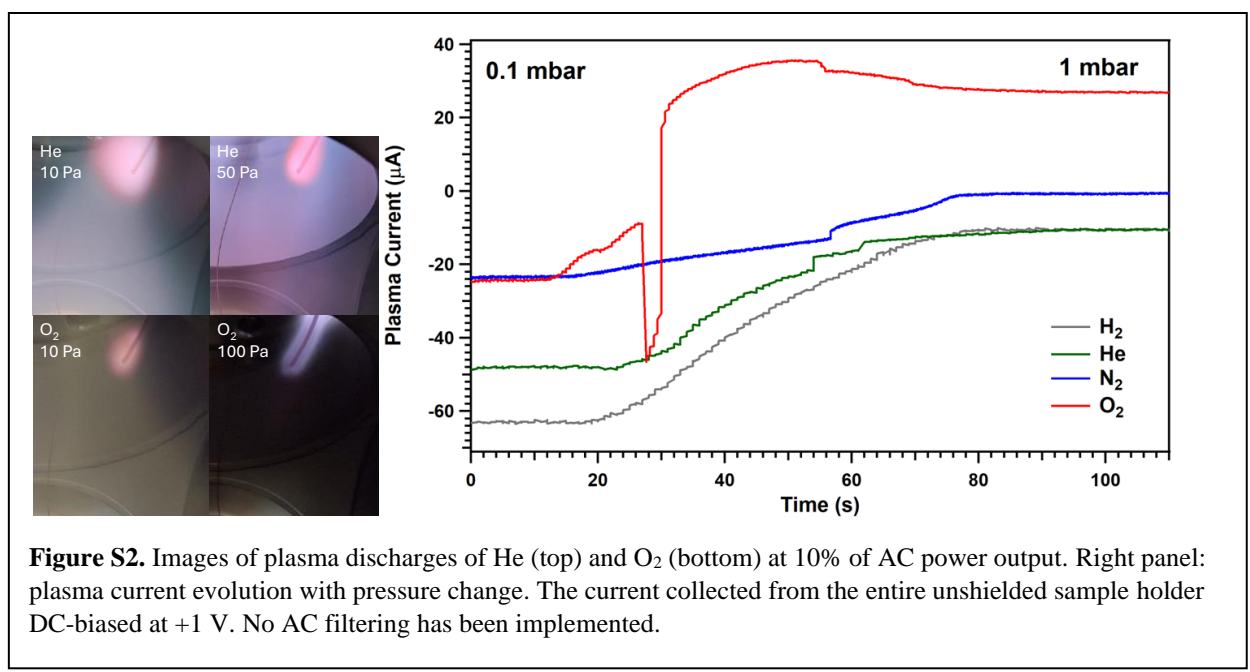

**Figure S2.** Images of plasma discharges of He (top) and O$_2$ (bottom) at 10% of AC power output. Right panel: plasma current evolution with pressure change. The current collected from the entire unshielded sample holder DC-biased at +1 V. No AC filtering has been implemented.

## C. Plasma-induced current is proportional to dial power settings

We found that the collected plasma-induced current is proportional to power dial settings and strongly

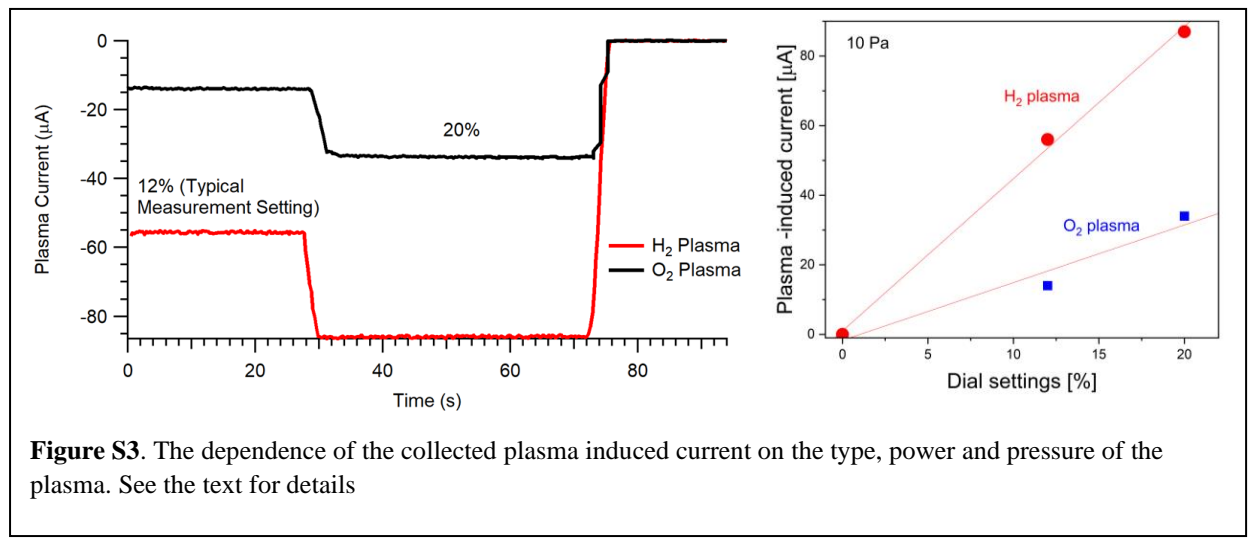

**Figure S3**. The dependence of the collected plasma induced current on the type, power and pressure of the plasma. See the text for details

depends on the gas used (see figure S3).

## D. Electrical characterization of the plasma source

Electrical plasma characterization was conducted by operating the copper wire electrode in a separate chamber (Fig. S4) equipped with standard plasma monitor instruments:

*Langmuir probe*: Characteristic current-voltage curves of plasma discharges are collected using an electrostatic tantalum wire of 10 mm in length and 0.4 mm in diameter. The probe tip can work in conjunction with a compensation electrode to provide accurate measurements of the plasma potential. The probe can be moved along one axis of the cross-chamber.

*Retarding field energy analyzer (RFEA)*: The ion flux and ion energy distribution from plasma discharges can be measured by applying retarding potentials to the set of parallel grids in a compact sensor (less than one mm-thick). The RFEA can be moved along the same axis as the Langmuir probe.

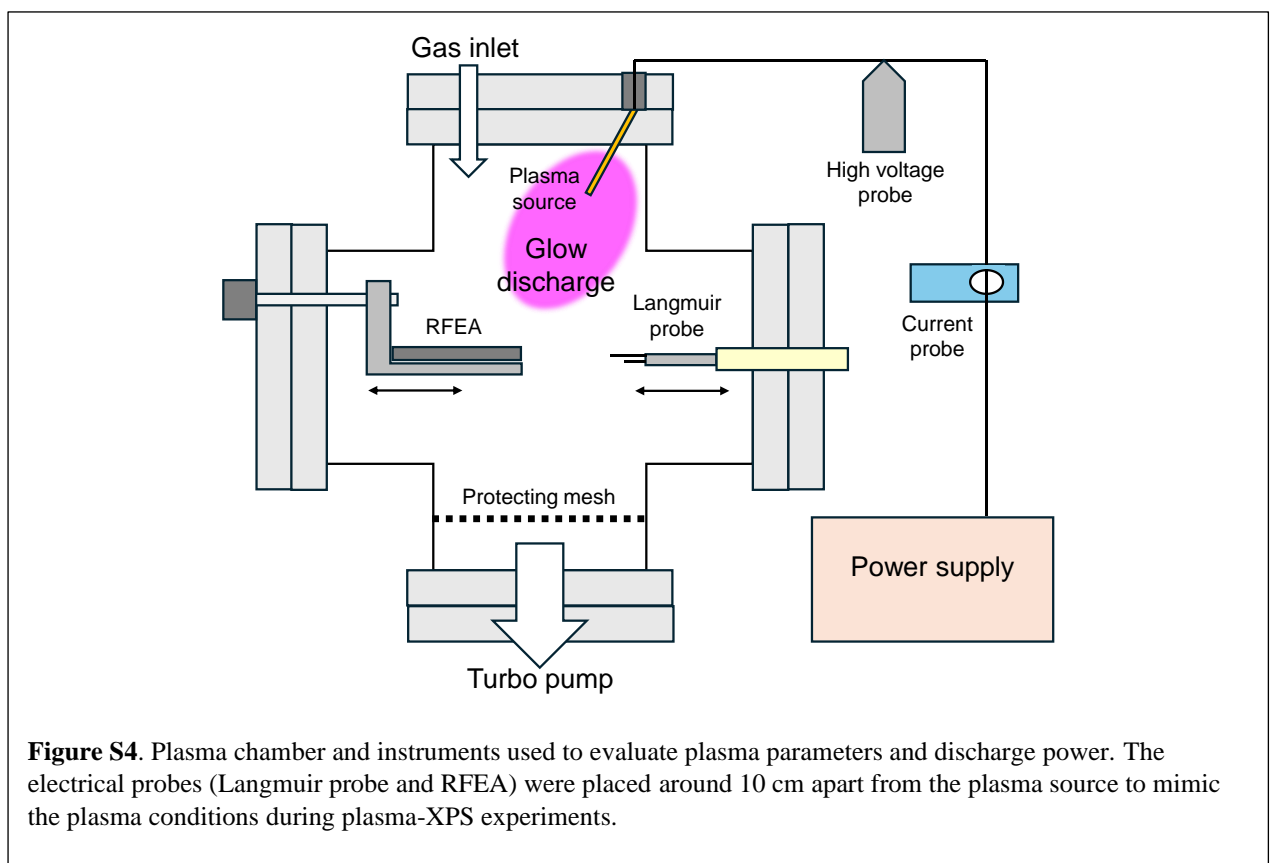

**Figure S4**. Plasma chamber and instruments used to evaluate plasma parameters and discharge power. The electrical probes (Langmuir probe and RFEA) were placed around 10 cm apart from the plasma source to mimic the plasma conditions during plasma-XPS experiments.

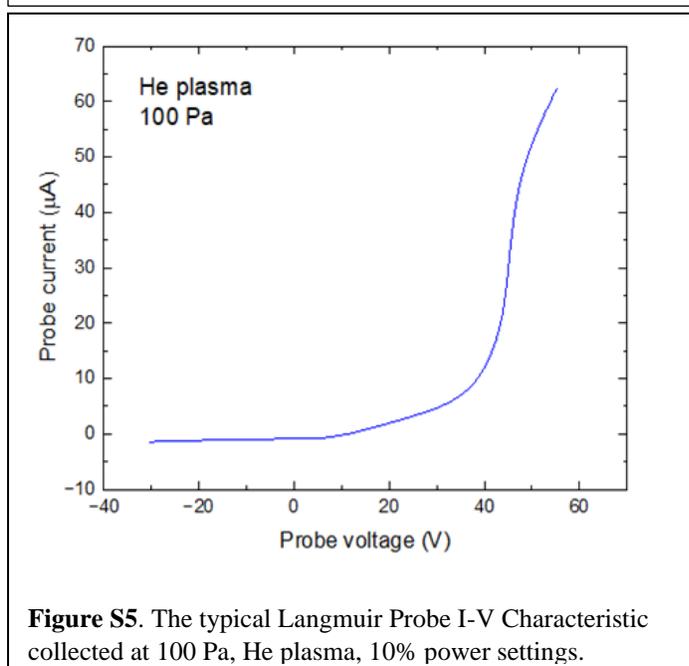

**Figure S5**. The typical Langmuir Probe I-V Characteristic collected at 100 Pa, He plasma, 10% power settings.

The main plasma parameters were measured using a Langmuir probe placed 10 cm away from the plasma source, roughly the same distance held between the plasma source and the sample in the plasma-XPS experiments. Fig. S5 shows a characteristic curve. The plasma potential varies between approximately 40V and 60V for all gases, pressures, and power settings used (10% - 40%). For all gases, the measured ion density dropped from approximately $10^{14}$ m$^{-3}$ at 10 Pa to around $10^{13}$ m$^{-3}$ as pressure increased to 100 Pa. Such a significant reduction is consistent with plasma constriction and plasma-induced current reduction as pressure increases (see Figs. S2, S3).

The ion flux density registered 10 cm away from the plasma source was below the sensitivity ($\approx 0.1$ A/m$^2$) of the used retarding field energy analyzer (RFEA) and could not be recorded.

## E. Discharge power

The electrical power consumed by the plasma discharges of Ar, He, $O_2$, $N_2$, and $H_2$, has been evaluated from measurements of discharge voltage waveforms, $V_{\text{dis}}(t)$, and discharge current waveforms, $I_{\text{dis}}(t)$, for a power output of 10 % set in the AC power supply (Fig. S6).

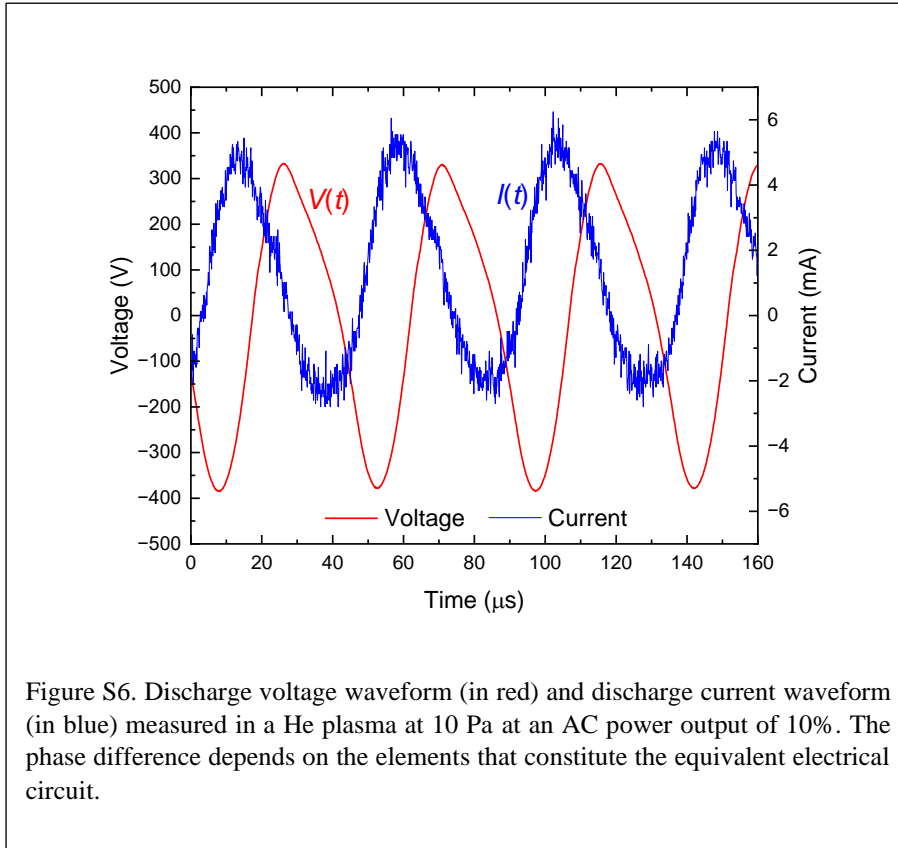

Figure S6. Discharge voltage waveform (in red) and discharge current waveform (in blue) measured in a He plasma at 10 Pa at an AC power output of 10%. The phase difference depends on the elements that constitute the equivalent electrical circuit.

The discharge current waveform is obtained by subtracting the current waveforms corresponding to plasma *on* and plasma *off* conditions. $I_{\text{dis}}(t) = I_{\text{on}}(t) - I_{\text{off}}(t)$.[1] Discharge power has been defined as the integral of instantaneous power, $V_{\text{dis}}(t)I_{\text{dis}}(t)$, averaged over one AC cycle with period $\tau$:[2]

$$\langle P \rangle = \frac{1}{\tau} \int_0^\tau V_{\text{dis}}(t)[I_{\text{on}}(t) - I_{\text{off}}(t)]\mathrm{d}t$$

The calculated discharge power values span between 0.1 W and 1 W and are summarized in Fig. S7.

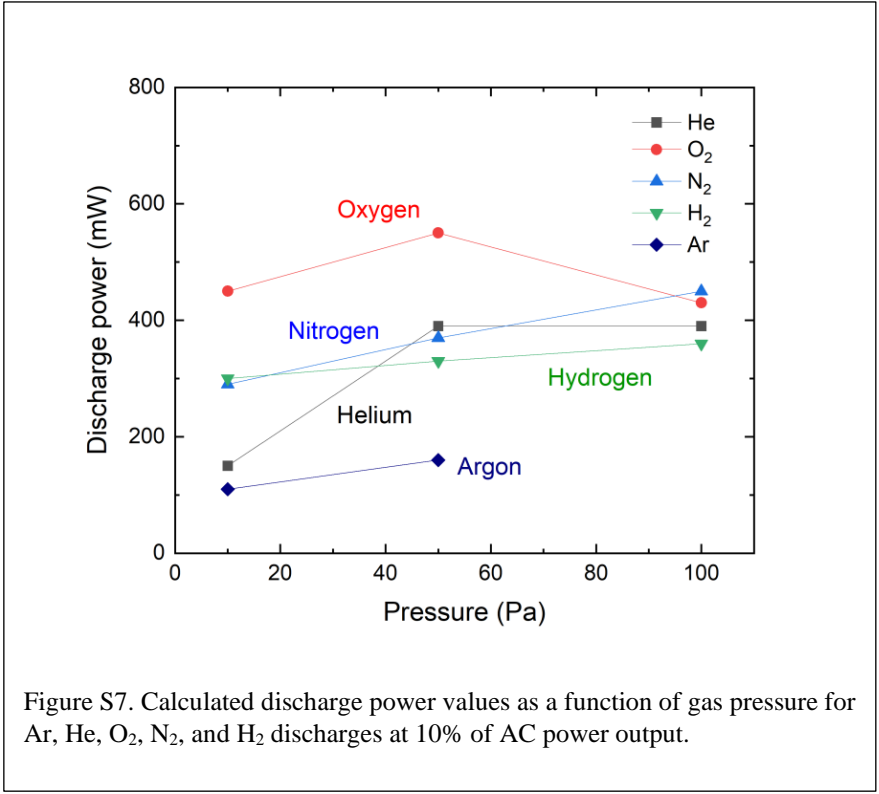

Figure S7. Calculated discharge power values as a function of gas pressure for Ar, He, O₂, N₂, and H₂ discharges at 10% of AC power output.

## F. Plasma vs AC frequencies consideration

First, we evaluate the time evolution of our AC discharge and estimate whether the near electrode sheaths can follow the changes of the applied voltage. The characteristic time scale of the sheath reconfiguration is controlled by the motion of ions, and therefore can be estimated based on the ion plasma frequency:

$$2\pi f_{pi} = \sqrt{\frac{n_i e^2 Z^2}{\varepsilon_0 M_i}}$$

where $f_{pi}$ is the ion plasma frequency, $n_i$ is the ion number density, $m_i$ is the ion mass, $Z$ is the ion charge number, $e$ is the electron charge, and $\varepsilon_0$ is the vacuum permittivity. The ion plasma frequency needs to be compared with the frequency $f$ of the applied voltage 22 kHz. Assuming the experimental values of ion number density, measured by the Langmuir probe, to be in the order of $n_i \approx 1 \times 10^{13} \ m^{-3}$, and the slowest ions in our experiments, molecular oxygen ion $O_2^+$ with $M_i = 32$ amu, we obtain for $f_{pi} \sim$ 120 kHz. This value is almost one order of magnitude higher compared to driving frequency $f$, and we may safely conclude that the electrode sheaths follow the changes of the applied voltage. Therefore, at the first approximation, our plasma can be treated as a plasma of a DC glow discharge, where the plasma-generating electrode acts either as a cathode or an anode.

## G. Plasma Oxidation of Au Fitting Parameters

All Au peaks were fit using the XPS fitting software [3]. The Au 4f were fit with an asymmetric Lorentzian (LA(1,1.1,120)) to account for the slight asymmetric tail of the 4f shape that overlaps with the oxide peak.

The unoxidized and annealed Au 4f was fit to this shape first to ensure a good and consistent fit. The metallic Au$^0$ 4f7/2 peak is positioned at 84.3 eV, with the 4f5/2 peak constrained to 3.67 eV higher energy and 0.75 times the area of the 4f7/2 peak. These fits were then further constrained to hold both position and full-width half maximum (FWHM) constant for fitting the oxidized spectrum. The oxide peaks were fit with mixed Gaussian Lorentzian (GL 95) following the energy positions provided from Tsai et al. [2]. The oxide peaks were also constrained to an energy separation of 3.67 eV and spin orbit split of 4:3. The best fit for the Au O 4f7/2 were at a photon energy of 85.6 eV, which agrees well with the 85.5 eV seen in multiple papers [4-5]. A slight variation in the Au-O energy position is expected as the oxide thickness (or oxide content at the surface) influences this energy [5], most likely due to a change in the vacuum level dependent on the semiconducting nature of the gold oxide.